\documentclass[12pt]{article}
\usepackage{graphicx,amssymb,epsfig,epsf} 
\usepackage{ulem}
\usepackage[usenames]{color}
\usepackage{rotating}
\usepackage{epsfig}
\usepackage{amsmath}
\usepackage{pstricks}
\usepackage{color}

\def\bt{\begin{table}}
\def\et{\end{table}}
\def\bc{\begin{center}}
 \def\ec{\end{center}}
\def\bi{\begin{itemize}}
\def\ei{\end{itemize}}
\def\bea{\begin{eqnarray}}
\def\eea{\end{eqnarray}}
\def\beas{\begin{eqnarray*}}
\def\eeas{\end{eqnarray*}}

\def\beq{\begin{equation}}
\def\eeq{\end{equation}}

\def\Journal#1#2#3#4{{#1} {\bf #2}, #3 (#4)}

\def\NPB{Nucl. Phys. B}

\def\PLB{{Phys. Lett.} B}

\def\PRL{Phys. Rev. Lett.}

\def\PRD{Phys. Rev. D}

\def\JHEP{JHEP}

\def\EPJ{Euro. Phys. J. C}

\def\IJMP{Int. J. Mod. Phys. A}

\def\JPG{J. Phys. G}

\def\Erratum{Erratum-ibid}

\begin{document}
\begin{flushright}
   {\bf OSU-HEP-14-06}\\
\end{flushright}

\vskip 30pt

\begin{center}

{\large A model for Dirac neutrino mass matrix with only four parameters}\\
\vskip 20pt
{Shreyashi Chakdar\footnote{chakdar@okstate.edu}, {Kirtiman Ghosh\footnote{kirti.gh@gmail.com}}} and S. Nandi\footnote{s.nandi@okstate.edu}   \\
\vskip 10pt
{Department of Physics and Oklahoma Center for High Energy Physics,\\
Oklahoma State University, Stillwater, OK 74078-3072, USA.}\\

\end{center}

\vskip 20pt

\abstract
{The exchange symmetry between  the muon neutrino and the tau neutrino for the neutrino mass matrix
has been very useful in understanding  the near maximal atmospheric neutrino mixing angle. However,
this symmetry can not be imposed at the Lagrangian level, since the charged lepton partners, muon and tau
do not satisfy this symmetry. We extend the Standard model to include three right handed singlet neutrinos,
and impose the most general symmetry between $\nu_{\mu R} $ and $\nu_{\tau R}$ sectors followed by a CP transformation of the leptonic sector at the Lagrangian level. This symmetry does not affect the charged leptons. With the additional assumption of the hermiticity of the ensuing Dirac neutrino mass matrix, we 
get a 4 parameter neutrino mass matrix in good agreement with the available neutrino data for the inverted neutrino mass hierarchy. The model also predicts the values of the three neutrino masses, and the leptonic CP violating phase which can be tested in the upcoming neutrino experiments.}


\vskip 30pt
\newpage
The mixing between different neutrino flavors was first hinted by the deficit of solar neutrino flux as measured in Earth. The solar neutrino deficit can be explained if we assume non-zero neutrino masses, mixings and hence, oscillation between different neutrino flavors. During last two decades different experiments on 
atmospheric ($\nu_\mu$ and $\bar\nu_\mu$) neutrinos (Super-K \cite{Wendell:2013kxa}, K2K \cite{Mariani:2008zz}, MINOS \cite{Barr:2013wta}), solar ($\nu_e$) neutrinos ( SNO \cite{Aharmim:2011vm}, Super-K \cite{Koshio:2013dta} , KamLAND \cite{Mitsui:2011zz}) as well as  reactor/accelerator ($\bar\nu_e$/$\nu_\mu$) neutrinos (Daya Bay \cite{An:2012eh}, RENO \cite{Ahn:2012nd}, Double Chooz \cite{Abe:2011fz}, T2K\cite{Abe:2011sj}, NO$\nu$a \cite{Patterson:2012zs}) provided us convincing evidences for non-zero neutrino masses and mixings.  All currently available data on the oscillations can be described assuming 3-flavor ($\nu_e,~\nu_\mu~{\rm and}~\nu_\tau$) neutrino mixing in vacuum. In the basis where the weak interaction is flavor diagonal and universal, the mass eigenstates ($\nu_1,~\nu_2~{\rm and}~\nu_3$) are related to the weak (flavors) eigenstates ($\nu_e,~\nu_\mu~{\rm and}~\nu_\tau$) as follows,
 \begin{equation} 
\left(\begin{matrix} \nu_e\\ \nu_\mu\\ \nu_\tau \end{matrix}\right) = U \left(\begin{matrix} \nu_1\\ \nu_2\\ \nu_3 \end{matrix}\right),
\end{equation}
where, $U$ is the $3\times 3$ neutrino mixing matrix \cite{Pontecorvo:1957cp,Maki:1962mu}. For Dirac neutrinos the mixing matrix $U$ can be parametrized by 3 angles ($\theta_{12}$, $\theta_{13}$ and $\theta_{23}$) and one CP violating phase ($\delta$): 
\begin{equation} 
U
= \left(\begin{matrix}
c_{12}c_{13} & s_{12}c_{13} & s_{13}e^{-i\delta} \\
-s_{12}c_{23}-c_{12}s_{23}s_{13}e^{i\delta} & c_{12}c_{23}-s_{12}s_{23}s_{13}e^{i\delta} & s_{23}c_{13} \\
s_{12}s_{23}-c_{12}c_{23}s_{13}e^{i\delta} & -c_{12}s_{23}-s_{12}c_{23}s_{13}e^{i\delta} & c_{23}c_{13}
\end{matrix}
\right),  
\label{eqn2}
\end{equation}
where, $c_{ij}={\rm Cos}\theta_{ij}$ and $s_{ij}={\rm Sin}\theta_{ij}$. A global analysis of neutrino oscillations data from different experiments give the best fit values for the three mixing angles and two squared-mass differences \cite{PDG}, $\Delta m_{ij}^2=m_i^2-m_j^2$. However there are several important  parameters yet to be measured. These include the value of the CP phase(s)  which will determine the magnitude of CP violation in the leptonic sector and the sign of $\Delta m^2_{32}$ which will determine whether the neutrino mass hierarchy is normal or inverted. Moreover, we also don't know yet if the neutrinos are Majorana or Dirac particles. 

The best fit values of the three mixing angles and two squared-mass differences along with their $3\sigma$ allowed range are presented in Table~\ref{mixing_data}. The experimental data in Table~\ref{mixing_data} shows two important properties: (i) There is a ${\cal O}(10^{2})$ hierarchy in the  squared-mass differences and (ii) The atmospheric and solar mixing angles ($\theta_{12}$ and $\theta_{23}$) are large whereas the reactor mixing angle ($\theta_{13}$) is very small. It is well known that the presence of tiny quantities or hierarchies indicates towards a protection symmetry in underlying scenario \cite{Hooft}. Example of one such well studied symmetry, in the context of neutrino physics, is the invariance of flavor neutrino mass matrix under interchange of $\nu_\mu$ and $\nu_\tau$ \cite{Nishiura,mu-tau,mu-tau0,mu-tau1,mu-tau2,CGN}.
It is easy to see from Eq.~\ref{eqn2} that the exact $\mu$-$\tau$ symmetry of the neutrino mixing matrix demands $s^2_{23}=0.5$ and $s_{13}=0$. Table~\ref{mixing_data} shows that $s^2_{23}=0.5$ is still within the $3\sigma$ of the central value however,  $s_{13}=0$ is already ruled out with more than $5\sigma$ C.L. Moreover, the charged leptons and left handed neutrinos are in the $SU(2)_L$ doublets and thus, the $\mu$-$\tau$ symmetry respected by the neutrinos should be respected by the charged leptons. However, the charged leptons clearly violate these symmetries at the Lagrangian level. Therefore, one can only impose $\mu$-$\tau$ symmetry as a symmetry of neutrino mass matrix not as a symmetry of the Lagrangian. This fact apparently disfavors the requirement of the $\mu$-$\tau$ symmetry.
\begin{table}[t]
\begin{center}
\begin{tabular}{|c|c|c|}
\hline
\hline
Parameter & best-fit ($\pm\sigma$) & $3\sigma$\\
\hline
$\Delta m^2_{21}[10^{-5} eV^2]$ & $7.53_{-0.22}^{+0.26}$ & 6.99 - 8.18 \\
\hline
$\Delta m^2[10^{-3} eV^2]$ & $2.43_{-0.10}^{+0.06}$ ($2.42_{-0.11}^{+0.07}$) & 2.19(2.17) - 2.62(2.61)\\
\hline
$\sin^2\theta_{12}$ & $0.307_{-0.016}^{+0.018}$ & 0.259 - 0.359 \\
\hline
$\sin^2\theta_{23}$ & $0.386_{-0.021}^{+0.024}$($0.392_{-0.022}^{+0.039}$) & 0.331(0.335) - 0.637(0.663)\\
\hline
$\sin^2\theta_{13}$ & $0.0241\pm 0.0025$($0.0244_{-0.0025}^{+0.0023}$) & 0.0169(0.0171) - 0.0313(0.0315)\\
\hline
\end {tabular}
\end{center}
\caption {The best-fit values and $3\sigma$ allowed ranges of the 3-neutrino oscillation parameters. The values (values in brackets) correspond to normal neutrino mass hierarchy (NH) i.e., $m_1<m_2<m_3$ (inverted neutrino mass hierarchy (IH) i.e., $m_3<m_1<m_2$). The definition of $\Delta m^2$ used is $\Delta m^2 = m_{3}^2 - (m_{2}^2 + m_{1}^2)/2$. Thus $\Delta m^2 = \Delta m_{31}^2 - m_{21}^2/2$ if $m_1 < m_2 < m_3$ and $\Delta m^2 = \Delta m_{32}^2 + m_{21}^2/2$ for $m_3 < m_1 < m_2$.}
\label{mixing_data}
\end{table}
\\

In this work, we have enlarged the SM field content by introducing three right handed $SU(2)_L$ singlet neutrino fields ($\nu_{eR},~\nu_{\mu R}$ and $\nu_{\tau R}$). We have also considered Yukawa terms for the neutrinos in order to give them Dirac masses\footnote{If the neutrinos get mass via the Yukawa couplings with the SH Higgs then the order of the neutrino Yukawa coupling should be about $10^{-12}$. However, there are interesting studies in the literature \cite{Gabriel:2006ns} which assume a discrete $Z_2$ symmetry and a second Higgs doublet with vacuum expectation value in the eV to keV range, in order to generate sub eV scale Dirac type neutrino masses with a Yukawa coupling of the order of charged lepton Yukawa coupling.}. In this frame work, we can demand a invariance of flavor neutrino mass terms under the interchange of the right handed muon neutrino ($\nu_{\mu R}$) and tau neutrino ($\nu_{\tau R}$).  The RH charged leptons and neutrinos are singlet under $SU(2)_L$ and thus they do not form a multiplate. Therefore, we can invoke any symmetry in the RH neutrino sector without inposing that symmetry in the charged lepton sector. If any symmetry  exists in the Dirac neutrino mass matrix under interchange of $\nu_{\mu R}$-$\nu_{\tau R}$ then this will be symmetry of the whole Lagrangian. We have constructed the different Dirac neutrino mass matrices assuming different kinds of symmetries in the $\nu_{\mu R}$ and $\nu_{\tau R}$ sector and tried to fit the experimentally observed quantities. Finally, we end up with a four parameter Dirac neutrino mass matrix which is based on the assumption of the Hermiticity\footnote{It is important to note that the assumption of Hermiticity is somewhat ad hoc i.e., Hermiticity of neutrino mass matrix is not an outcome of symmetry argument. However, we have shown in the following that with this assumption, the existing neutrino data can completely determine the mass matrix for the Dirac neutrinos with particular predictions for the neutrino masses and the CP violating phase which can be tested at the ongoing and future neutrino experiments. Therefore, in our analysis, the assumption of hermiticity of neutrino mass matrix is a purely phenomenological assumption. However, in the future, there might be some compelling theoretical framework which requires the hermiticity of neutrino mass matrix.} of the Dirac neutrino mass matrix and a particular symmetry between  $\nu_{\mu R}$ and $\nu_{\tau R}$. We have also shown that assuming IH in the neutrino sector, this four parameter neutrino mass matrix is consistent with the observed values of the three mixing angles and two squared-mass differences listed in Table~\ref{mixing_data}, and also makes  definite
predictions for the values of the three neutrino masses and the leptonic CP violating phase .

The most general Dirac neutrino mass matrix contain 9 complex parameters and can be written as: 
\begin{equation} 
 \bold {M}_{\nu}= \left(\begin{matrix}
m_{e_L e_R} & m_{e_L \mu_R} & m_{e_L \tau_R} \\
m_{\mu_L e_R} & m_{\mu_L \mu_R} & m_{\mu_L \tau_R} \\
m_{\tau_L e_R} & m_{\tau_L \mu_R} & m_{e_L \tau_R} \\
\end{matrix}
\right). 
\label{general_matrix}
\end{equation}
On this 18 parameter Dirac neutrino mass matrix, we have imposed the following conditions:
\begin{itemize}
\item We have assumed the hermiticity of the neutrino mass matrix. As a result of this assumption, the diagonal elements of Eq.~\ref{general_matrix} become real and off-diagonal elements become complex conjugate of each other: $m_{\mu_L e_R}=m_{e_L \mu_R}^*$, $m_{\tau_L e_R}=m_{e_L \tau_R}^*$ and $m_{\tau_L \mu_R}=m_{\mu_L \tau_R}^*$. Therefore, after demanding the hermiticity, we have a 9 parameter neutrino mass matrix. 
%
\end{itemize}

The hermitian neutrino mass matrix is given in the flavor basis by
\begin{equation} 
M_{\nu} = U_{\nu} \bold {M}_{\nu}^{diag} U_{\nu}^{\dagger},
\label{eqn0}
\end{equation}
where, ${M}_{\nu}^{diag}$ is the diagonal neutrino mass matrix in the mass basis. Two squared-mass differences of the neutrinos are known from the experiments. Therefore, ${M}_{\nu}^{diag}$ can be constructed with only one mass as unknown. For IH, the diagonal neutrino mass matrix is given by,
\begin{equation} 
 \bold {M}_{\nu}^{diag}= \left(\begin{matrix}
\sqrt{m_3^2 + 0.002315} & 0 & 0 \\
0 & \sqrt{m_3^2 + 0.00239} & 0\\
0 & 0 & m_3
\end{matrix}
\right),  
\end{equation}
where, $m_3$ is the unknown mass and we have used the central values of the squared-mass differences listed in Table~\ref{mixing_data} for IH. In the mixing matrix $U$, there are three angles and one phase. The mixing angles are already measured (see Table~\ref{mixing_data} for their central values and $3\sigma$ range) with good precision. In our analysis, we have considered the IH central values for the $s_{12}^2$ and $s_{13}^2$. However, we have considered $s_{23}^2=0.5$ which is not the central value but well within $3\sigma$ of the central value. 

If we assume one particular neutrino mass hierarchy, there are still two quantities unknown in for the Dirac neutrinos namely, the mass $m_3$ in the diagonal mass matrix and the CP violating phase ($\delta$) in the mixing matrix. In our analysis, we have scanned unknown parameters ($m_3$ and $\delta$) over a range of values and tried to find out a constrainted phenomenological neutrino mass matrix which is consistent with the 5 experimental results (three mixing angles and two squared-mass differences). 
Our phenomenological results are summarized in the following:
\begin{figure}[t]
\begin{center}
\includegraphics[angle =-90, width=0.7\textwidth]{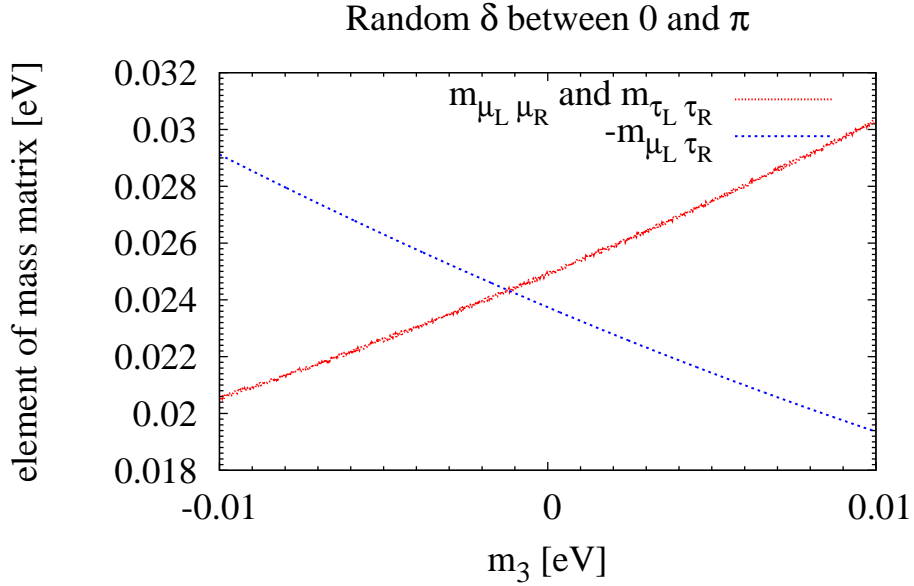}
\end{center}
\caption{The elements $m_{\mu_L \mu_R}$ and real part of -$m_{\mu_L \tau_R}$ of the Dirac neutrino mass matrix in Eq.~\ref{general_matrix} as a function of $m_3$. The other free parameter $\delta$ was randomly varied between 0 and $\pi$. We have used IH central values for the $\Delta m_{21}^2$, $\Delta m^2$, $s^2_{12}$ and $s^{2}_{13}$ from Table~\ref{mixing_data} and for $s^2_{23}$, we choose $s^2_{23}=0.5$.} 
\label{xm}
\end{figure}

\begin{figure}[t]
\begin{center}
\includegraphics[angle =-90, width=0.49\textwidth]{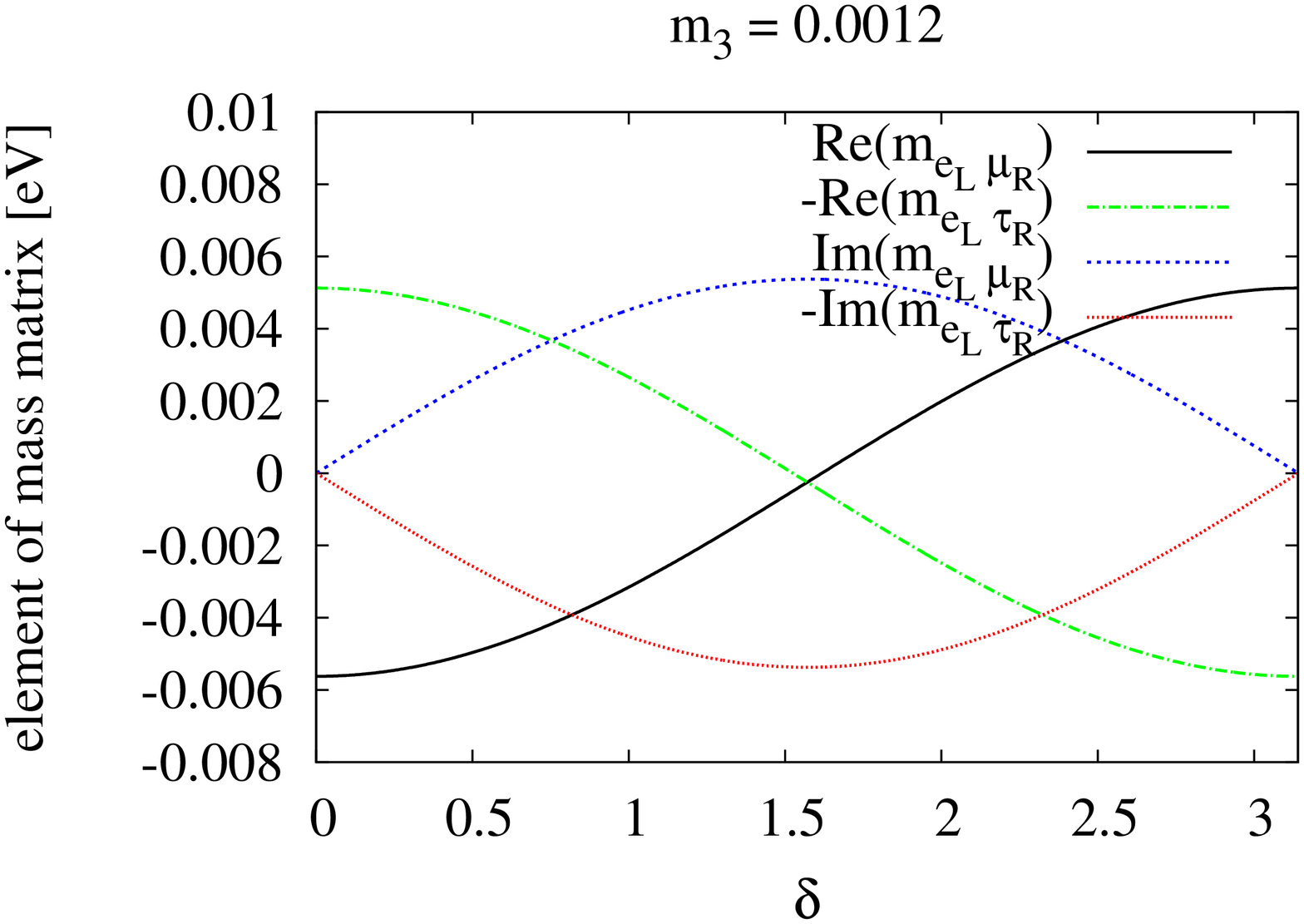}
\includegraphics[angle =-90, width=0.49\textwidth]{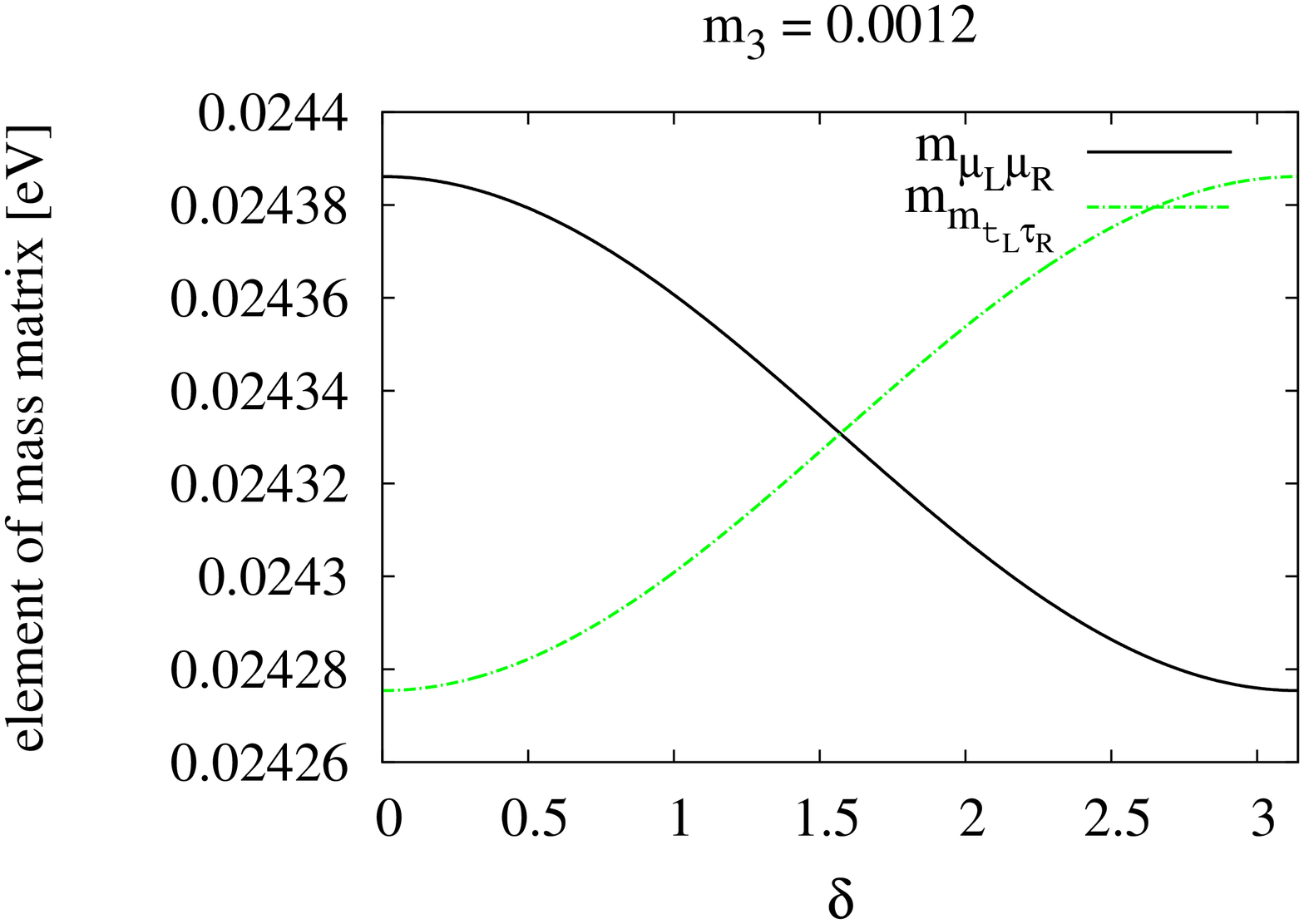}
\end{center}
\caption{Left panel: The real and imaginary part of the elements $m_{e_L \mu_R}$ and $m_{e_L \tau_R}$ of the Dirac neutrino mass matrix in Eq.~\ref{general_matrix} as a function of $\delta$ (in radian) for $m_3=-1.198 \times 10^{-3}$. Right panel: The diagonal elements $m_{\mu_L\mu_R}$ and $m_{\tau_L \tau_R}$ of the Dirac neutrino mass matrix in Eq.~\ref{general_matrix} as a function of $\delta$ for $m_3=-1.198 \times 10^{-3}$. We have used IH central values for the $\Delta m_{21}^2$, $\Delta m^2$, $s^2_{12}$ and $s^{2}_{13}$ from Table~\ref{mixing_data} and for $s^2_{23}$, we choose $s^2_{23}=0.5$.} 
\label{del}
\end{figure}

\begin{itemize}
\item In Fig.~\ref{xm}, we have presented  $m_{\mu_L \mu_R}$ and real part of -$m_{\mu_L \tau_R}$ elements of the Dirac neutrino mass matrix in Eq.~\ref{general_matrix} as a function of $m_3$. The other free parameter $\delta$ was randomly varied between 0 and $\pi$. Fig.~\ref{xm} shows that two curves interests each other at $m_3=-1.198 \times 10^{-3}$eV. 
\item In Fig.~\ref{del}, we have presented real and imaginary parts of the elements $m_{e_L \mu_R}$ and $m_{e_L \tau_R}$ (left panel) and diagonal elements $m_{\mu_L\mu_R}$ and $m_{\tau_L \tau_R}$ (right panel) of the Dirac neutrino mass matrix in Eq.~\ref{general_matrix} as a function of $\delta$ for $m_3=-1.198 \times 10^{-3}$ eV. Fig.~\ref{del} shows that 
a constrained neutrino mass matrix is obtained for $\delta=\pi/2$ and $m_3=-1.198 \times 10^{-3}$ eV. The numerical form of the mass matrix in the falvour basis for $\delta=\pi/2$ and $m_3=-1.198 \times 10^{-3}$ eV is given by,
\end{itemize}
\begin{equation} 
\left(\begin{matrix}
4.72\times 10^{-2} & 2.49 \times 10^{-4}-5.37 \times 10^{-3}i & -2.49 \times 10^{-4}-5.37 \times 10^{-3}i \\
 2.49 \times 10^{-4}+5.37 \times 10^{-3}i & 2.43 \times 10^{-2}  & -2.43 \times 10^{-2} \\
-2.49 \times 10^{-4}+5.37 \times 10^{-3}i  & -2.43 \times 10^{-2} 
&  2.43 \times 10^{-2} \\
\end{matrix}
\right), 
\label{num_matrix}
\end{equation} 
It is important to note that the mass matrix in Eq.~\ref{num_matrix} is a four parameter matrix can be written as,
\begin{equation}
 \bold {M}_{\nu}^{pheno}=
\left(\begin{matrix}
a & b e^{i\eta} & -b e^{-i\eta} \\
 b e^{-i\eta} & c & -c \\
-b e^{i\eta}  & -c &  c \\
\end{matrix}
\right), 
\label{param_matrix}
\end{equation} 
with $a=4.72\times 10^{-2}$, $b=5.38\times 10^{-3}$, $c=2.43 \times 10^{-2}$ and $\eta=272.6^{0}$. We  now  search for  symmetry in the $\nu_{\mu R}$-$\nu_{\tau R}$ sector which is consistent with the structure of the phenomenological neutrino mass matrix in Eq.~\ref{param_matrix}. 

The most general transformation in the $\nu_{\mu R}$-$\nu_{\tau R}$ sector can be written as,
\begin{equation} 
\Psi_{R}=\left(\begin{matrix}
\nu_e\\
\nu_\mu\\
\nu_\tau\\
\end{matrix}
\right) \to
\left(\begin{matrix}
e^{i\phi_1} & 0 & 0 \\
0 & p e^{i\phi_2} & -q e^{-i\phi_3}\\
0 & q e^{i\phi_3} & p e^{-i\phi_2}\\
\end{matrix}
\right)~\Psi_{R} \to U_{R}\Psi_R, 
\label{r_trans}
\end{equation} 
where, $p^2+q^2=1$ and $\phi_1,~\phi_2$ and $\phi_3$ are the arbitrary phases. As already discussed in the beginning of this paper, we do not want to introduce any symmetry in $\nu_{\mu L}$-$\nu_{\tau L}$ sector in order to make the symmetry as the symmetry of the Lagrangian. However, phase transformation for the left-handed neutrino fields are still allowed:
 \begin{equation} 
\Psi_{L}=\left(\begin{matrix}
\nu_e\\
\nu_\mu\\
\nu_\tau\\
\end{matrix}
\right) \to
\left(\begin{matrix}
e^{-i\theta_1} & 0 & 0 \\
0 & e^{-i\theta_2} & 0\\
0 & 0 & e^{-i\theta_3}\\
\end{matrix}
\right)~\Psi_{L}\to U_{L}\Psi_L, 
\label{l_trans}
\end{equation}  
We have demanded the invariance under simultaneous transformations $\Psi_R \rightarrow  U_R \Psi_R$ and $\Psi_L \to U_L \Psi_L$ followed by a complex conjugation of the couplings. Complex conjugation of the couplings is equivalent to making a CP transformation. In the rest of this article, the symmetry  under above mentioned transformations followed by a  CP transformation is denoted as {\it $\nu_{\mu R}$-$\nu_{\tau R}$ reflection} symmetry. As a consequence of the {\it $\nu_{\mu R}$-$\nu_{\tau R}$ reflection} symmetry, we obtain the following matrix equation:
\begin{equation} 
\left [ U_L^\dagger  \bold {M}_{\nu}^{pheno} U_R \right]^*~=~\bold {M}_{\nu}^{pheno}.
\label{meq}
\end{equation}  
The most general solution of Eq.~\ref{meq} is given by
\begin{eqnarray}
\phi_1=n_1\pi - {\rm cos}^{-1}\left[(-1)^{n_2} p\right ]~&;&~\theta_1=n_1\pi + {\rm cos}^{-1}\left[(-1)^{n_2} p\right ];\nonumber\\
\phi_2=n_2\pi~&;&~\theta_2={\rm cos}^{-1}\left[(-1)^{n_2} p\right ];\nonumber\\
\phi_2=\left(n_3+\frac{1}{2}\right)\pi~&;&~\theta_2={\rm sin}^{-1}\left[(-1)^{n_3} q\right ]
\end{eqnarray}
and
\begin{equation}
\eta~=~\frac{n\pi}{2};
\end{equation}
where, $n,~n_1,~n_2$ and $n_3$ are arbitrary integers. The trivial solution ($n_1=0,~n_2=0$ and $n_3=0$) of Eq.~\ref{meq} physically corresponds to  a symmetry under interchange of $\nu_{\mu R} \leftrightarrow -i\nu_{\tau R}$ followed by a CP transformation with $\eta=0^0,~90^0,~180^0,~270^0,....$. However, the phenomenological neutrino mass matrix under consideration (Eq.~\ref{num_matrix} and Eq.~\ref{param_matrix}) corresponds to $\eta=272.6^0$. Therefore, tiny violation of the symmetry under interchange of $\nu_{\mu R} \leftrightarrow -i\nu_{\tau R}$ followed by a CP transformation is required to satisfy all the experimental results. 

To summarize, we have considered Dirac neutrino mass matrix and investigated the possible symmetries in the $\nu_{\mu R}$-$\nu_{\tau R}$ sector. In order to ensure that the imposed condition is a symmetry of the Lagrangian (not only the symmetry of the neutrino mass matrix in the flavor basis), we have restricted the requirements only to the singlet right-handed muon and tau neutrinos. Assuming the hermiticity of the neutrino mass matrix, we have obtained a particular structure of the phenomenological Dirac neutrino mass matrix with only 4 parameters. This 4 parameter Dirac neutrino mass matrix can explain all five (two squared-mass differences and three mixing angles) experimental results in the neutrino sector with particular predictions for the absolute values of the neutrino masses ($m_1=4.81\times 10^{-2},~m_2=4.89\times 10^{-2}~{\rm and}~m_3=-1.198\times 10^{-3}$ eV) and CP violating phase $\delta=270^{0}$. We have shown that the 4 parameters phenomenological mass matrix corresponds to a symmetry under interchange of $\nu_{\mu R} \leftrightarrow -i\nu_{\tau R}$ followed by a CP transformation  with a tiny violation of this symmetry to accomodate a value of the phase $\delta=272.6^{0}$ as required by the mass matrix in Eq. (6).  \\
\noindent{\bf Acknowledgment:}  The work of SC, KG and SN was supported in part by the U.S. Department of Energy Grant Number DE-SC0010108.


\begin{thebibliography}{99}
\bibitem{Wendell:2013kxa} 
  R.~Wendell [Super-Kamiokande Collaboration],
  Nucl.\ Phys.\ Proc.\ Suppl.\  {\bf 237-238}, 163 (2013).


\bibitem{Mariani:2008zz} 
  C.~Mariani [K2K Collaboration],
  AIP Conf.\ Proc.\  {\bf 981}, 247 (2008)

\bibitem{Barr:2013wta} 
  G.~Barr [MINOS Collaboration],
  PoS ICHEP {\bf 2012}, 398 (2013).

\bibitem{Aharmim:2011vm} 
  B.~Aharmim {\it et al.}  [SNO Collaboration],
  Phys.\ Rev.\ C {\bf 88}, 025501 (2013).

\bibitem{Koshio:2013dta} 
  Y.~Koshio [Super-Kamiokande Collaboration],
  PoS ICHEP {\bf 2012}, 371 (2013).

\bibitem{Mitsui:2011zz} 
  T.~Mitsui [KamLAND Collaboration],
  Nucl.\ Phys.\ Proc.\ Suppl.\  {\bf 221}, 193 (2011).

\bibitem{An:2012eh} 
  F.~P.~An {\it et al.}  [DAYA-BAY Collaboration],
  Phys.\ Rev.\ Lett.\  {\bf 108}, 171803 (2012).


\bibitem{Ahn:2012nd} 
  J.~K.~Ahn {\it et al.}  [RENO Collaboration],
  Phys.\ Rev.\ Lett.\  {\bf 108}, 191802 (2012).

\bibitem{Abe:2011fz} 
  Y.~Abe {\it et al.}  [DOUBLE-CHOOZ Collaboration],
  Phys.\ Rev.\ Lett.\  {\bf 108}, 131801 (2012).
  
\bibitem{Abe:2011sj} 
  K.~Abe {\it et al.}  [T2K Collaboration],
  Phys.\ Rev.\ Lett.\  {\bf 107}, 041801 (2011)
  [arXiv:1106.2822 [hep-ex]].
  
\bibitem{Patterson:2012zs} 
  R.~B.~Patterson [NOvA Collaboration],
  Nucl.\ Phys.\ Proc.\ Suppl.\  {\bf 235-236}, 151 (2013).

\bibitem{Pontecorvo:1957cp} 
  B.~Pontecorvo,
  Sov.\ Phys.\ JETP {\bf 6}, 429 (1957)
  [Zh.\ Eksp.\ Teor.\ Fiz.\  {\bf 33}, 549 (1957)],
  B.~Pontecorvo,
  Sov.\ Phys.\ JETP {\bf 7}, 172 (1958)
  [Zh.\ Eksp.\ Teor.\ Fiz.\  {\bf 34}, 247 (1957)].
\bibitem{Maki:1962mu} 
  Z.~Maki, M.~Nakagawa and S.~Sakata,
  Prog.\ Theor.\ Phys.\  {\bf 28}, 870 (1962).
\bibitem{PDG}
J.~Beringer {\it et al.}~(Particle Data~Group).
\newblock {\em Phys. Rev.}, {\bf D86}:117702, 2012.


  \bibitem{Hooft}  
  G. 't Hooft, in {\it Recent Development in Gauge Theories: Proceedings of the 
Cargese Summer Institute} edited by G. 't Hooft et al. (Plenum Press, New York, 
1980), p.135 (NATO Advanced Study Institutes Series: Series B, Physics, Vol 59).

\bibitem{Nishiura}
	T. Fukuyama and H. Nishiura, in {\it Proceedings of International Workshop on 
Masses and Mixings of Quarks and Leptons} edited by Y. Koide (World Scientific, 
Singapore, 1997), p.252; ``Mass Matrix of Majorana Neutrinos", [arXive:hep-ph
/9702253];
	Y. Koide, H. Nishiura, K. Matsuda, T. Kikuchi and T. Fukuyama, \Journal{\PRD}{66}
{093006}{2002};
	Y. Koide, \Journal{\PRD}{69}{093001}{2004};
	K. Matsuda and H. Nishiura, \Journal{\PRD}{69}{117302}{2004}; \Journal{\PRD}{71}
{073001}{2005}; \Journal{\PRD}{72}{033011}{2005}; \Journal{\PRD}{73}{013008}{2006}.

\bibitem{mu-tau}
	R.N. Mohapatra and S. Nussinov,\Journal{\PRD}{60}{013002}{1999};
	C.S. Lam, \Journal{\PLB}{507}{214}{2001}; \Journal{\PRD}{71}{093001}{2005};
	E. Ma and M. Raidal, \Journal{\PRL}{87}{011802}{2001}; [\Journal{\Erratum}{87}
{159901}{2001}];
	A. Datta and P.J. O'Donnell, \Journal{\PRD}{72}{113002}{2005}.

\bibitem{mu-tau0}
	T. Kitabayashi and M. Yasu\`{e}, \Journal{\PLB}{524}{308}{2002}; 
	\Journal{\IJMP}{17}{2519}{2002}; \Journal{\PRD}{67}{015006}{2003};
	I. Aizawa, M. Ishiguro, T. Kitabayashi and M. Yasu\`{e}, \Journal{\PRD}{70}{015011}
{2004};
	I. Aizawa, T. Kitabayashi and M. Yasu\`{e}, \Journal{\PRD}{71}{075011}{2005}.

\bibitem{mu-tau1}
	W. Grimus and L. Lavoura,  \Journal{\JHEP}{0107}{045}{2001}; \Journal{\EPJ}{28}
{123}{2003}; 
	\Journal{\PLB}{572}{189}{2003}; \Journal{\PLB}{579}{113}{2004}; \Journal{\JPG}
{30}{1073}{2004}; 
	\Journal{\JHEP}{0508}{013}{2005}; \Journal{\JHEP}{0508}{013}{2005};	
	W. Grimus, A.S. Joshipura, S. Kaneko, L. Lavoura and M. Tanimoto, \Journal{\JHEP}
{0407}{078}{2004}; 
	W. Grimus, A.S. Joshipura, S. Kaneko, L. Lavoura, H. Sawanaka and M. Tanimoto, 
\Journal{\NPB}{713}{151}{2005}; 
	M. Tanimoto, ``Prediction of $U_{e3}$ and $\cos 2\theta_{23}$ from Discrete 
Symmetry",  to be published in {\it Proceedings of XXXXth Rencontres de Moriond:
 Electroweak Interactions and Unified Theories}, La Thuile, Italy (March 5-12, 
2005), [arXive:hep-ph/0505031].

\bibitem{mu-tau2}
	R.N. Mohapatra,  \Journal{\JHEP}{0410}{027}{2004};
	R.N. Mohapatra and S. Nasri,  \Journal{\PRD}{71}{033001}{2005};
	R.N. Mohapatra, S. Nasri and Hai-Bo Yu, \Journal{\PLB}{615}{231}{2005}; \Journal
{\PRD}{72}{033007}{2005}.


\bibitem{CGN} 
  S.~Chakdar, K.~Ghosh and S.~Nandi,
  arXiv:1403.1544 [hep-ph].

\bibitem{Gabriel:2006ns} 
  S.~Gabriel and S.~Nandi,
  Phys.\ Lett.\ B {\bf 655}, 141 (2007)
  [hep-ph/0610253]. 
 

\end{thebibliography}
\end{document}